\documentclass[a4paper,10pt]{article}
\usepackage[utf8]{inputenc}
\usepackage[a4paper, total={6in, 10in}]{geometry}
\usepackage{amsmath}
\usepackage{amssymb}  
\usepackage{graphicx}
\usepackage{float}
\usepackage{array}
\usepackage{listings}
\usepackage{xcolor}
\usepackage{hyperref}
\usepackage[nameinlink]{cleveref}

\usepackage{cite}
\usepackage{subcaption}
\usepackage{tikz}
\usepackage{multicol}
\usepackage{multirow}
\DeclareUnicodeCharacter{2212}{\textendash}

\begin{document}
\title{Elasto-viscous regime in coalescence of viscoelastic droplets}


\author{Pallavi Katre$^{1}$, Bimalendu Mahapatra$^{1}$, Manaswita Karmakar$^{1}$, Sarang Jagdish$^{2}$,\\
Navin Kumar Chandra$^{1}$, and Aloke Kumar$^{1}$\footnote{alokekumar@iisc.ac.in} \\
\\
$^{1}$Department of Mechanical Engineering, IISc Bangalore, India-560012\\
$^{2}$Department of Mechanical Engineering, IIT Indore, India-453552
}
 
\date{} 
\maketitle

\section*{Abstract}
We report a regime transition in the coalescence of concentrated polymeric droplets in a pendant-pendant configuration. While Newtonian droplet coalescence has been extensively studied with distinct identification of viscous and inertial regimes, the presence of polymers introduces additional regimes governed by elasticity and molecular relaxation effects. The coalescence process is typically characterized by the neck radius, $R$, of the liquid bridge connecting the two droplets, following a power-law relation with time: $R=at^{b}$. Most of the existing studies, including Newtonian and non-Newtonian fluids, report a unique value of $b$ for a given fluid. In contrast, our findings reveal that elasticity induces a temporal transition from one $b$ values to another, marking a shift in the coalescence regime. In particular, our measured 
$b$ value falls in the sub-Newtonian regime, highlighting the role of elasticity in governing the dynamics. We conducted two-dimensional simulations using a volume-of-fluid framework with the exponential Phan-Thien–Tanner model, which quantitatively reproduced Newtonian benchmarks and accurately captured viscoelasticity induced neck growth in close agreement with experiments. Furthermore, we determined the curvature experimentally, as the assumptions typically employed in the literature to approximate axial curvature are not universally valid.

\section*{Introduction}

Coalescence \cite{eggers1999coalescence,aarts2005hydrodynamics,xia2017vortex}of liquid droplets is central to many important processes, from natural events like raindrop formation \cite{villermaux2009single} in clouds to industrial applications such as emulsion stability \cite{cho2003creating, paik2003rapid}, combustion\cite{orme1997experiments}, inkjet printing \cite{duineveld2003stability, ashgriz2011handbook}, spray painting, coating \cite{ashgriz1990coalescence} and even biological phenomena like cellular aggregation \cite{oriola2022arrested, ambrose2015mediated, grosser2021cell, ongenae2021activity}. Dynamics of fluid flow in droplet coalescence is observed in the form of the formation and growth of a liquid bridge. 
The temporal evolution of the neck region during droplet coalescence in Newtonian fluids is governed by a balance of viscous forces, inertial forces, and Laplace pressure \cite{hopper1984coalescence, hopper1990plane, wu2004scaling}. Depending on the fluid’s viscosity, the growth of the neck radius $R$ with time $t$ follows either a viscous or inertial regime, described by well-established scaling laws \cite{eggers1999coalescence, paulsen2013approach}, where \( R \sim t \) scaling is applicable in the viscous regime, and \( R \sim t^{1/2} \) is applicable for the inertial regime in the coalescence of freely suspended drops. In contrast, polymeric fluids introduce elasticity into this balance, which significantly alters the coalescence dynamics. Elastic stresses, particularly near the neck where polymer chains are highly stretched, lead to deviations from classical Newtonian behavior \cite{dekker2022elasticity}, necessitating a modified framework to capture the viscoelastic response. A recent study by Varma et al. \cite{varma2020universality} reported a scaling of \( R \sim t^{0.36} \)
for polymeric fluids and demonstrated a noticeable delay in the coalescence process, attributed to the influence of elastic effects. Their findings indicate that even with an increase in polymer concentration up to 20 times the critical concentration, the coalescence power-law exponent ($b$) remains invariant \cite{varma2022rheocoalescence}. For values of 
$b$ in the range $0<b<0.5$, the coalescence is characterized as being in the sub-Newtonian regime \cite{rajput2023sub, sudheer2025sub}. The sub-Newtonian regime is significant because it occurs across various classes of complex fluids \cite{sudheer2025sub}, though the precise mechanism is not yet understood. It signifies the onset of elasticity in polymeric fluids and the emergence of non-Newtonian effects, offering key insights into the underlying molecular and rheological mechanisms governing coalescence dynamics. The theoretical limit at $b=0$ corresponds to the arrested coalescence state. Arrested coalescence refers to the incomplete merging of droplets where coalescence is halted due to elasticity, interfacial forces, or particle jamming at the interface \cite{pawar2012arrested, oriola2022arrested}. These different regimes, as determined by varying values of $b$, are illustrated in Figure \ref{fig:fig1}a for a better understanding of these regimes.

Although the scaling exponent $b$ characterizes the observed coalescence regimes, the fundamental driving mechanism remains the Laplace pressure which depends on two principal curvatures, radial ($1/R$) and axial curvature ($k = 1/R_k$). These curvatures are of similar magnitude (Figure \ref{fig:fig1}c), so both must be considered in the momentum conservation equation in the radial direction. Although axial curvature drives the coalescence via a negative Laplace pressure jump, $R_k$ is difficult to measure experimentally. In contrast, $R$ can be accurately tracked, so most studies report its variation. While both R and $R_k$ appear in theoretical models, $R_k$ is often eliminated using simplifying assumptions for comparison with experiments using geometric self-similarity as \cite{chandra2025yoga,xia2019universality},
\begin{equation}
\frac{R_k}{R} \;\approx\; \frac{R}{2R_0} \;\ll\; 1
\end{equation}
Here, 
$R_0$ is the initial droplet radius. This assumption holds only for the inertial-limited-viscous regime and fails to capture the viscoelastic behavior \cite{paulsen2012inexorable}. The one-dimensional models valid for Newtonian fluids break down in viscoelastic cases, where flow becomes inherently multidimensional due to coupled viscous and elastic effects, especially near the neck where polymer chains stretch significantly \cite{oratis2023coalescence}. The neck shape depends on both viscosity and elasticity, sharpening with increasing viscosity and polymer concentration as chain stretching alters the stress singularity during coalescence \cite{thoroddsen2005coalescence, dekker2022elasticity, bouillant2022rapid}.

This complex interplay between elasticity and viscosity can give rise to multiple coalescence regimes. However, most of the
previous studies on droplet coalescence typically reported a single regime, likely due to their short observation times \cite{varma2022rheocoalescence}. Rostami et al. \cite{rostami2025coalescence} showed that polymer addition induces two distinct regimes in sessile–sessile coalescence, evident from neck-width evolution. Similarly, Chen and Yong \cite{chen2022viscoelastic} numerically identified viscous and viscoelastic regimes in attractive microgels at very early stages, beyond experimental reach, though without examining power-law exponents. Despite extensive experimental \cite{yao2005coalescence, aarts2005hydrodynamics, burton2007role, yokota2011dimensional, kumar2020coalescence, singh2024newtonian,sudheer2025sub} and numerical \cite{thompson2012inviscid, martinez1995viscous, sprittles2012coalescence} efforts, key aspects of coalescence remain unresolved. Here, by extending experiments to longer timescales and higher $c/c^*$, we report for the first time a regime transition in the neck radius in pendant–pendant coalescence.


We study droplet coalescence in polyethylene oxide (PEO) solutions at varying concentrations, comparing Newtonian and polymeric fluids. Experiments reveal two distinct regimes in high-viscosity polymeric fluids: an elasticity-dominated regime followed by a elasto-viscous regime where both viscous and elastic forces are comparable. Elasticity significantly alters the power-law exponent $b$ from Newtonian behavior, with its influence reflected in the axial curvature analysis. In parallel, we performed two-dimensional simulations using a volume-of-fluid framework using the exponential Phan-Thien–Tanner model~\cite{thien1977new,10.1122/1.549481}. These simulations not only reproduced the classical Newtonian viscous and inertial scalings but also captured the elasticity-induced slowdown of neck growth, in quantitative agreement with experiments. The combined experimental–numerical approach enables us to establish, for the first time, a clear transition between elastic and elasto-viscous regimes in pendant–pendant droplet coalescence. To enhance practical relevance, we also examine coalescence in a consumer-grade complex fluid, shampoo.


\section*{Results and discussion}

\begin{figure}[h]
\centering
\includegraphics[width=.9\linewidth]{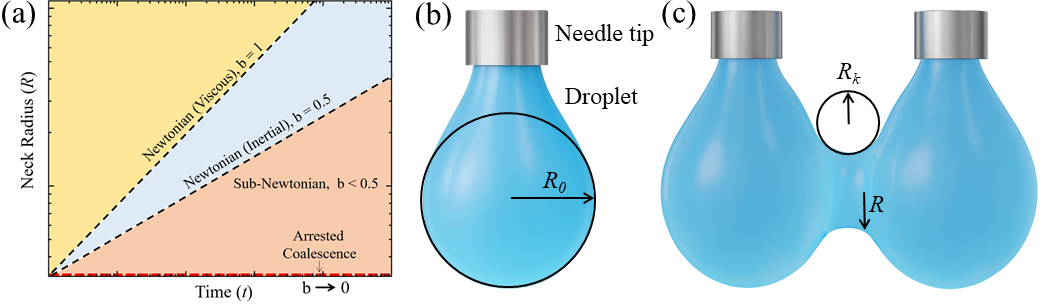}
\caption{(a) Power-law exponent corresponding to various coalescence regimes, (b) schematic of the pendent drop showing the calculation of initial droplet radius $R_0$, and (c) coalescence of droplets in pendent-pendent configuration showing the neck radius $R$ and radius of curvature $R_k$ in the axial direction.}
\label{fig:fig1}
\end{figure}

In the present study, two pendant droplets of similar size were brought into contact by bringing one droplet towards the other with an approach velocity of around $10^{-4}$ m/s. Upon contact, a liquid neck forms and subsequently expands until the system reaches the thermodynamic equilibrium of a single droplet. The neck growth, described by the radius $R$, is driven by Laplace pressure, which diverges at first contact due to infinite interfacial curvature. We investigated Newtonian and polymeric fluids with different Ohnesorge numbers ($Oh = \eta/\sqrt{\rho \gamma R_0}$), where $\eta$, $\rho$, and $\gamma$ denote zero-shear viscosity, density, and surface tension, respectively. Fluid properties and corresponding $Oh$ values are listed in Supplementary Table \ref{tab:T1}. Polymer concentration is expressed as $c/c^*$, with $c^*$ being the critical concentration. The initial droplet radius $R_0$ was obtained by fitting a circle  to the initial droplet shape (Figure \ref{fig:fig1}b). Coalescence dynamics were quantified by tracking the evolution of neck radius $R$. We also estimate the axial curvature ($k$) at the neck by fitting a 4th-order polynomial to the local profile, which is approximated as a circular segment (Figure \ref{fig:fig1}c). The curvature is then calculated from the second derivative at the neck, with interactive fitting employed to ensure accuracy.

\begin{figure}[t]
\centering
\includegraphics[width=.95\linewidth]{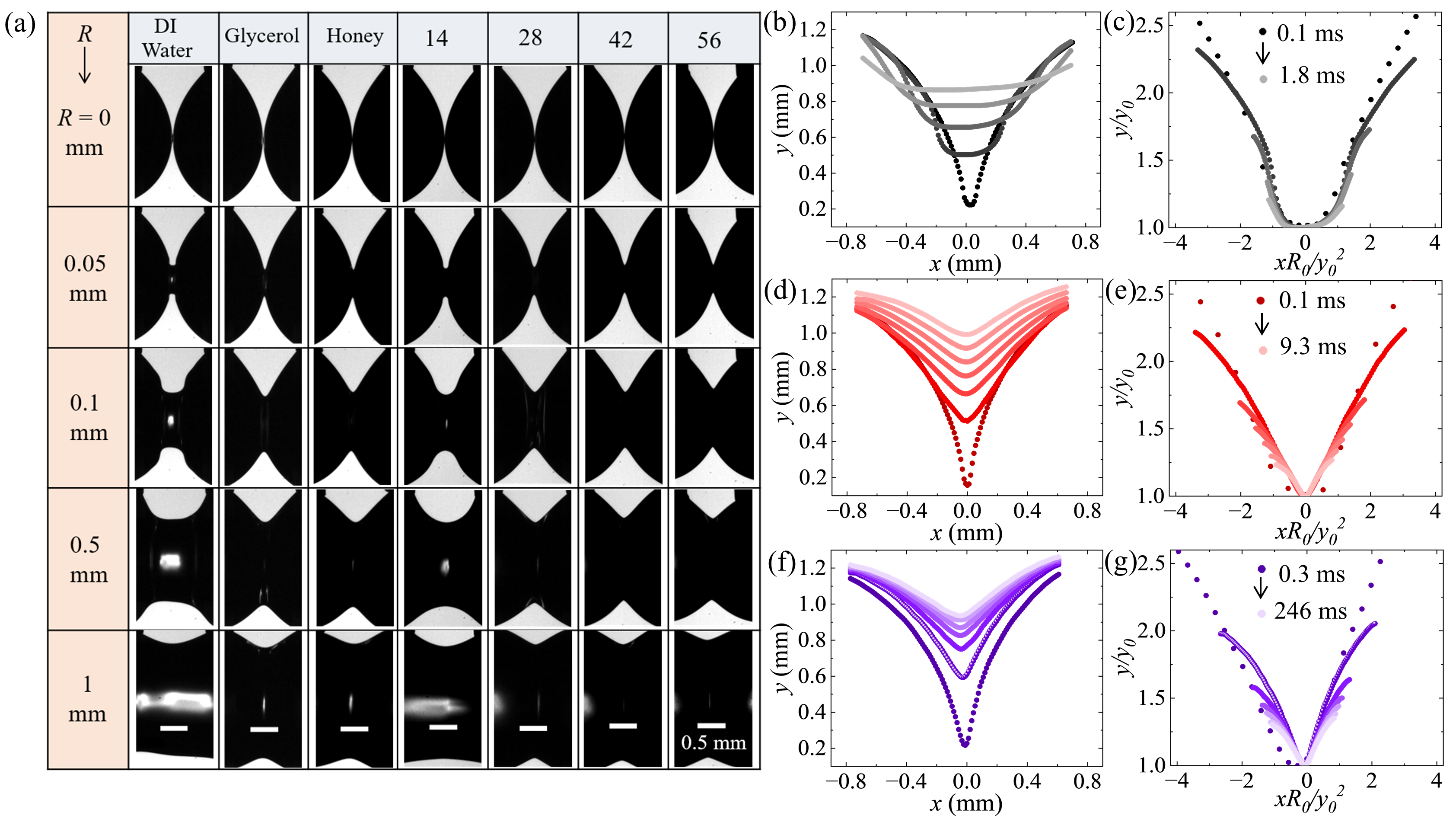}
\caption{(a) Snapshots of droplet shapes displaying varying neck radius for DI Water, Glycerol, Honey and polymer solutions with different $c/c^*$ of 14, 28, 42 and 56, (b) bridge profiles at different times and their corresponding rescaled profiles according to the Newtonian scaling laws for (b, c) DI Water, and polymer solution with (d,e) $c/c^*$ = 28 and (f,g) $c/c^*$ = 56.}
\label{fig:fig2}
\end{figure}

Figure \ref{fig:fig2}a shows snapshots of droplet shapes at equal neck radii for the fluids studied. Among Newtonian fluids, DI water exhibits flatter axial curvature, while glycerol and honey display sharper initial curvature that gradually flattens, consistent with the effect of higher viscosity \cite{thoroddsen2005coalescence}. For polymer solutions, $c/c^* = 14$ shows curvature similar to water, but at $c/c^* \geq 28$ the curvature becomes significantly sharper and remains so throughout coalescence, highlighting the role of both viscosity and elasticity \cite{kaneelil2025coalescenc, bouillant2022rapid}.

To enable a precise comparison, contours were extracted from these snapshots (Figures \ref{fig:fig2}b, d, f). For DI water (Figure \ref{fig:fig2}b), the curvature is relatively flat, but it sharpens progressively with increasing polymer concentration (Figures \ref{fig:fig2}d, f), reflecting growing elastic stresses from stretched polymer chains. These stresses disrupt the self-similar scaling observed typically in Newtonian coalescence \cite{dekker2022elasticity}, where the bridge profiles are rescaled using the standard similarity variables, $R_0/y_0^2$ and $1/y_0$ on horizontal and vertical axes, respectively. $y_0$ is taken at time t = 0. Indeed, DI water profiles collapse onto a master curve under Newtonian rescaling,  as shown in Figure \ref{fig:fig2}c, confirming self-similarity. By contrast, PEO solutions fail to collapse under the same scaling (Figures \ref{fig:fig2}e, g), demonstrating that elastic stresses dominate the local bridge dynamics, with deviations growing stronger at higher $c/c^*$.

\begin{figure}[t]
\centering
\includegraphics[width=1\linewidth]{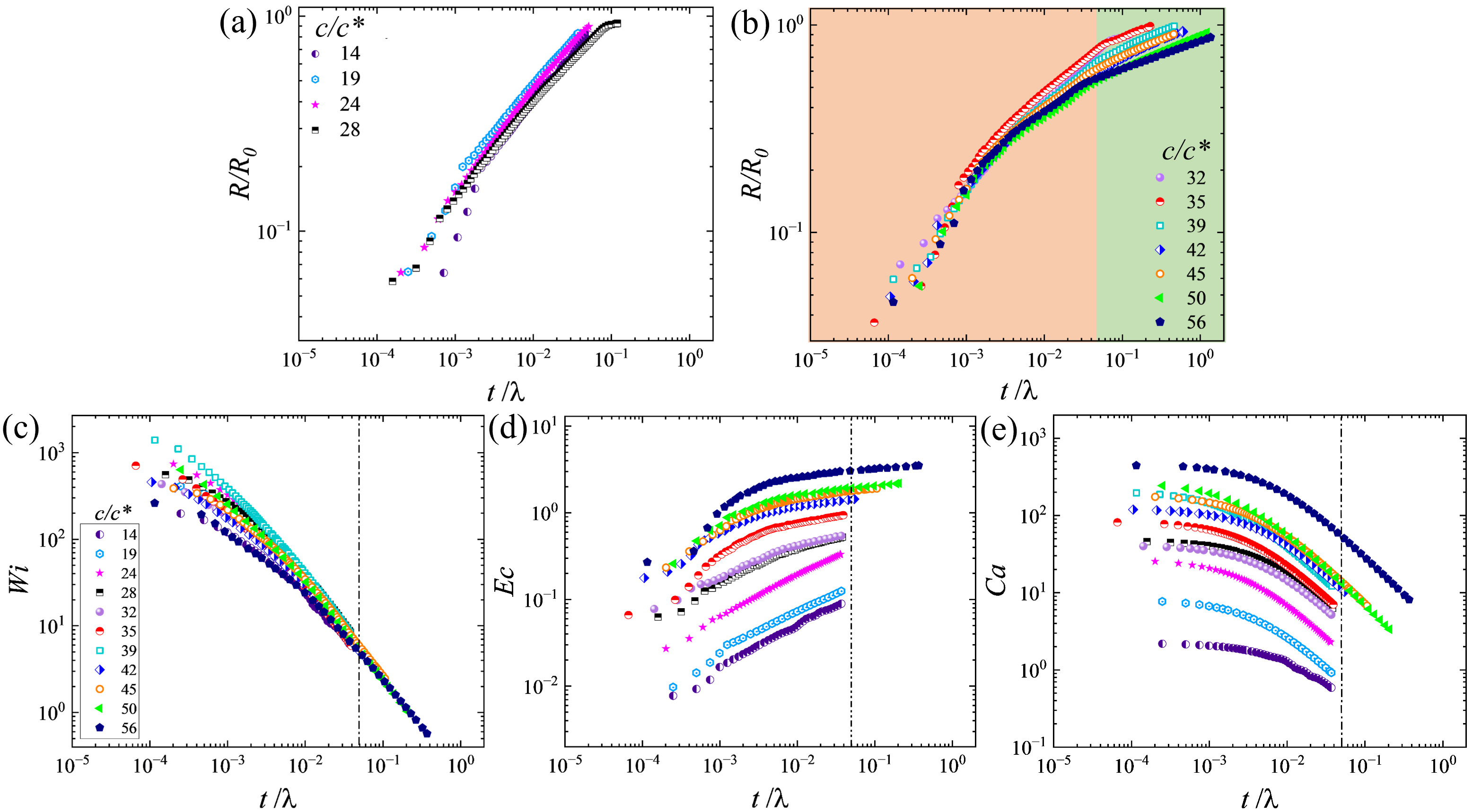}
\caption{Temporal evolution of neck radius for polymeric droplets with different $c/c^*$, (a) which do not exhibit a transition ($c/c^* < 28$), and (b) which do show a transition $c/c^* > 28$ with normalised time.  Variation of dimensionless numbers (c) Weissenberg number, (d) Elastocapillary number, and (e) Capillary number with normalised time.}
\label{fig:fig3}
\end{figure}

From these droplet shapes, we quantify the neck radius of the bridge, and its temporal evolution is shown in Figure \ref{fig:fig3} (a,b). Time is normalised by the relaxation time of the polymer solution ($\lambda$), measured from CaBER-DoS experiments. For Newtonian fluids (Supplementary Figure \ref{fig:S1}) and polymer solutions with low $c/c^* \leq 28$ (Figure \ref{fig:fig3}a), the neck radius exhibits a single regime.  For $c/c^* = 28$, a new regime begins at the end of the coalescence process, but only for a very short time. However, for higher concentrations ($c/c^* >$ 28), We qualitatively observe two distinct regimes, distinguished by their slopes (Figure \ref{fig:fig3}b). These two regions are represented by the different colors, merely illustrates the transition for the data shown in the figure. 
In the early regime ($t^* = t/\lambda < 0.05$), the fluid's elastic response dominates because polymer chains get stretched as the coalescence process starts. These polymer chains require time to relax and begin contributing to viscous flow after deformation starts \cite{ferry1980viscoelastic, rostami2025coalescence}. When a polymer solution is first sheared, the polymers take time to respond and are initially not fully engaged in the flow, delaying their contribution to the solution’s viscosity \cite{vereroudakis2023repeated}.This time-dependent effect is more pronounced in solutions with higher $c/c^*$. In the later stage, viscous behaviour becomes comparable with the elastic effects, showing the presence of elasto-viscous regime.

Next, we examine how these regimes affect the power-law scaling of neck growth, $R \sim t^b$. For Newtonian fluids and polymer solutions with $c/c^*\leq28$, a single power-law fit describes the coalescence dynamics. At higher concentrations ($c/c^*>28$), however, two distinct regimes emerge, requiring separate fits with different exponents (Supplementary Table \ref{T2}). In Newtonian coalescence, increasing viscosity suppresses inertial effects, causing the viscous-dominated regime to emerge earlier and persist longer \cite{xia2019universality}. As a result, the measured exponent $b$ shifts toward the viscous limit of 1. For example, water gives 
$b\approx0.45$, consistent with inertial scaling \cite{ristenpart2006coalescence}, whereas highly viscous honey approaches 
$b\approx1$ as viscous scaling dominates \cite{eggers1999coalescence}. By contrast, polymer solutions ($c/c^*$ = 14–28) show 
$b$ decreasing from 0.46 to 0.38 despite higher viscosity, indicating that elastic stresses delay or suppress the onset of viscous scaling.
The exponent $b$ reflects coalescence dynamics rather than speed \cite{chandra2025yoga}: viscosity advances the onset of viscous scaling, pushing $b$ toward 1, whereas elasticity delays or suppresses this onset, lowering 
$b$ toward 0; in both cases, the coalescence process becomes slower. For $c/c^* \geq 32$, two exponents appear, with regime 2 consistently lower than regime 1. Across both regimes, 
$b$  decreases systematically with increasing 
$c/c^*$, highlighting the growing influence of elasticity. In the later elasto-viscous regime ($t^* > 0.05$), viscosity becomes important, yet $b$ does not approach unity because elastic stresses from stretched polymer chains counteract viscous scaling. This reveals the competing roles of viscosity and elasticity in non-Newtonian droplet dynamics. As the values of $b$ are below 0.5 for all polymeric droplets, they clearly lie within the sub-Newtonian regime.

\begin{table}[h!]
\caption{Approximate orders of magnitude of $Wi$, $Ec$, and $Ca$ for low and high $c/c^*$ droplets in early and late stages.}
\centering
\begin{tabular}{|c|c|c|c|}
\hline
Regime & $Wi = \frac{\lambda}{R} \frac{dR}{dt}$ & $Ec = \frac{\eta R}{\lambda \sigma}  $ & $Ca = \frac{\eta}{\sigma} \frac{dR}{dt}$ \\ \hline
Low $c/c^*$ & $\gg \mathcal{O}(1)$ & $\leq \mathcal{O}(1)$ & $\mathcal{O}(1\text{--}10)$ \\ \hline
High $c/c^* > 28$, early & $\gg \mathcal{O}(1)$ & $\mathcal{O}(1)$ & $\gg \mathcal{O}(1)$ \\ \hline
High $c/c^* > 28$, late & $\mathcal{O}(1)$ & $\mathcal{O}(1)$ & $\mathcal{O}(1\text{--}10)$ \\ \hline
\end{tabular}

\label{tab:dim_numbers}
\end{table}

These regime transitions and the trends in $b$ can be rationalized by order-of-magnitude estimates of the Weissenberg ($Wi$), Elastocapillary ($Ec$), and Capillary ($Ca$) numbers (Figure~\ref{fig:fig3}c–e, Table~\ref{tab:dim_numbers}). For low concentrations ($c/c^* < 28$, Figure~\ref{fig:fig3}a), the neck grows in a single, continuous regime dominated by elasticity. Although viscosity is relatively low, elastic stresses resist capillary-driven deformation, resulting in $Wi \gg \mathcal{O}(1)$ (Figure~\ref{fig:fig3}c). The Elastocapillary number $Ec \leq \mathcal{O}(1)$ (Figure~\ref{fig:fig3}d) indicates that elastic and capillary forces are comparable or slightly weaker than capillarity, while $Ca = \mathcal{O}(1-10)$ (Figure~\ref{fig:fig3}e) remains moderate. The weak viscous resistance prevents the emergence of a second regime, and the coalescence remains essentially elasticity-dominated throughout.  

For higher concentrations ($c/c^* > 28$, Figure~\ref{fig:fig3}b), both elasticity and viscosity increase, leading to two distinct regimes. In the early stage (orange-shaded region), elastic stresses dominate over viscous and capillary contributions, as indicated by $Wi \gg\mathcal{O}(1)$ (Figure~\ref{fig:fig3}c) and $Ec = \mathcal{O}(1)$ (Figure~\ref{fig:fig3}d), while $Ca \gg \mathcal{O}(1)$ (Figure~\ref{fig:fig3}e) shows that viscous effects are also significant. The predominance of elasticity in this regime reduces the neck growth exponent, yielding $R \sim (t)^b$ with $b < 0.5$. At later times (green-shaded region), elastic, viscous, and capillary stresses become comparable, with $Wi = \mathcal{O}(1)$, $Ec = \mathcal{O}(1)$, and $Ca = \mathcal{O}(1\text{--}10)$. This competition yields two effective exponents, with the later exponent always lower than the early one.

We also explain the instantaneous influence of elasticity during droplet coalescence by considering the strain-rate scales. Extensional rheometry of the polymer solutions provides the critical strain rate, $\dot{\varepsilon}_c$, above which elastic stresses become significant (Supplementary Figure \ref{fig:S3} a and b). During coalescence, the instantaneous strain rate at the neck exceeds $\dot{\varepsilon}_c$ by approximately two orders of magnitude, stretching the polymer chains immediately and generating elastic stresses almost instantaneously. Consequently, elasticity dominates from the very onset, reducing the neck growth exponent relative to Newtonian or low-viscosity regimes.


\begin{figure}[t]
\centering
\includegraphics[width=1\linewidth]{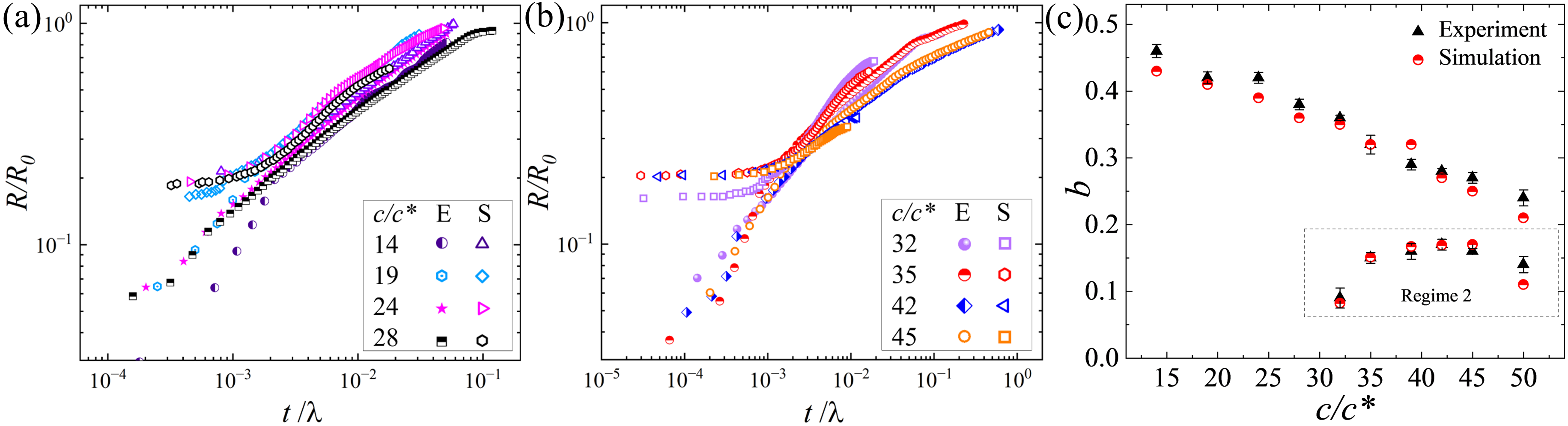}
    \caption{Comparison of experimental and numerical results for the coalescence of viscoelastic droplets. (a) Neck radius evolution in Regime~1 ($c/c^{*} = 14, 19, 24, 28$). (b) Neck radius evolution in Regime~2 ($c/c^{*} = 32, 35, 42, 45$). (c) Power-law exponents $b$ extracted from experimental and numerical data across all concentrations.}
\label{fig:fig6} 
\end{figure}

Having established the scaling trends experimentally, we first verified them against Newtonian benchmarks: water reproduced inertial scaling with $b \approx 0.5$, and honey approached viscous scaling with $b \approx 1$. For viscoelastic droplets, simulations using the exponential Phan–Thien–Tanner model~\cite{thien1977new,10.1122/1.549481} captured the reduction in $b$ caused by polymer elasticity, in close agreement with experiments (Figure~\ref{fig:fig6}). For low concentrations ($c/c^* < 28$), both experiments and simulations produced a single regime with $0.3 \leq b \leq 0.45$, consistent with elasticity-dominated dynamics ($Wi \gg \mathcal{O}(1)$, $Ec \leq \mathcal{O}(1)$). For higher concentrations, two regimes were consistently observed: in Regime 1, $b$ dropped to $0.1\text{--}0.2$ with strong elastic dominance ($Wi \gg \mathcal{O}(1)$, $Ca \gg \mathcal{O}(1)$), while in Regime 2, comparable elastic, viscous, and capillary stresses ($Wi \sim \mathcal{O}(1)$, $Ec \sim \mathcal{O}(1)$, $Ca \sim \mathcal{O}(1\text{--}10)$) yielded a lower exponent. The close agreement between experimental and numerical results validates our numerical framework as a reliable tool for modeling coalescence in viscoelastic fluids, where the interplay of elasticity, viscosity, and interfacial tension is crucial for determining scaling behavior. It is important to note that, although the polymer exhibits shear-thinning at high shear rates, our shear rheometry shows that the coalescence process operates well below the threshold required to activate measurable shear-thinning. Under these conditions, the exponential Phan–Thien–Tanner model reduces to a constant-viscosity viscoelastic description, and the close match between experimental and numerical neck dynamics demonstrates that shear-thinning plays a negligible role in droplet coalescence.

The preceding discussion centered on the neck radius and its transition during coalescence. We now examine the axial curvature ($k$), since the Laplace pressure driving coalescence depends on two principal curvatures: radial ($1/R$) and axial ($k$). While $R$ is readily measured, $R_k$ is difficult to estimate experimentally, even though it governs the negative Laplace jump. Both curvatures appear in the momentum conservation equation in radial direction \cite{xia2019universality}:  
\begin{align}
 \frac{1}{2} \rho U^2 - \sigma \left( \frac{1}{R_k} - \frac{1}{R} + \frac{2}{R_0} \right) - \eta \left( -\frac{\sqrt{\pi}\, U}{2 R_k} + O\left( \frac{U}{R} \right) \right) = 0,
\end{align}

where $U$ is the coalescence speed. Most prior studies remove $R_k$ using the geometric relation $\tfrac{R_k}{R} \approx \tfrac{R}{2R_0}$, though this assumption is not always valid. Our experiments show that $k$ depends on both $R$ and fluid rheology (Supplementary Figure \ref{fig:Supl-S2}c), and that $1/R$ and $1/R_k$ are of comparable magnitude, making it difficult to neglect either. The detailed analysis of the axial curvature is given in the supplementary section: Axial curvature (\ref{sec:axial_curv}).

 \begin{figure}[H]
\centering
\includegraphics[width=.5\linewidth]{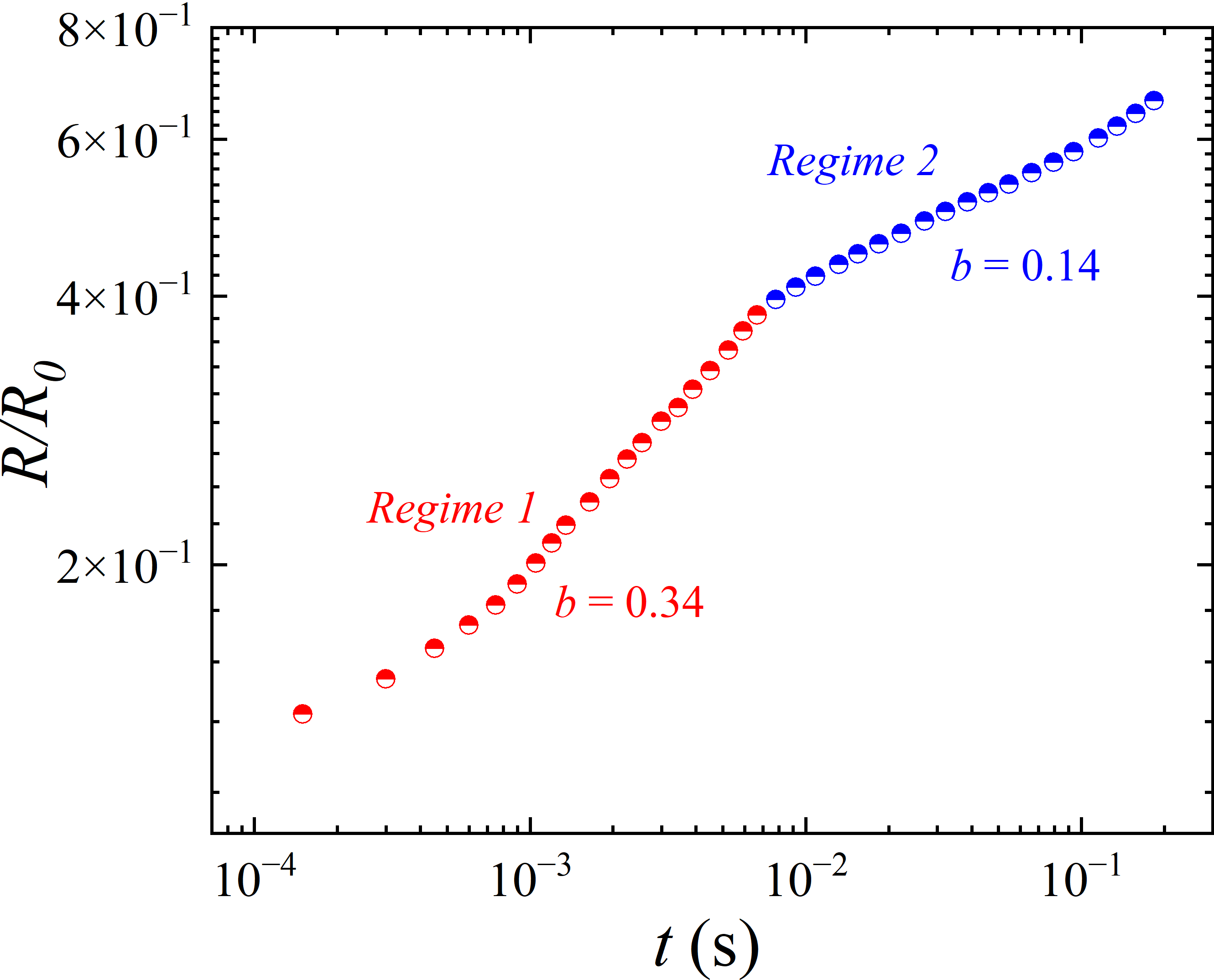}
\caption{Variation of normalized neck radius of shampoo (Dove) with time.}
\label{fig:fig4}
\end{figure}

To explore these observed regime transitions in daily used products, we conducted coalescence experiments using a common consumer product, namely shampoo. In this study, we used Dove shampoo (intense repair) as the test sample. We observed that the droplet coalescence behavior of shampoo closely resembles that of polymer solutions with high $c/c^*$. Two distinct regimes are seen in the coalescence process, as shown in Figure \ref{fig:fig4}. Additionally, even with the high viscosity of the shampoo (with a zero-shear viscosity of 18 Pa.s), the growth exponent is reduced to 0.34 in Regime 1 and 0.14 in Regime 2. This reduction highlights the significant influence of elasticity on the coalescence process of shampoo. The presence of multiple regimes and elasticity-driven deviations in such everyday products highlights the broader relevance and utility of our approach for characterizing coalescence behavior in practical, non-ideal systems.

\section*{Materials and methods}

In this study, several Newtonian and polymeric fluids with varying Ohnesorge numbers were examined. The Newtonian fluids, water, glycerol, and honey covered an $Oh$ range of 0.0033–14. Polyethylene oxide (PEO) with a molecular weight of 
$M_w = 4 \times 10^6$ gm/mol was used as the representative polymeric fluid, with $Oh$ values spanning 1.34–369. Eleven PEO solutions of different concentrations were prepared by dissolving PEO powder in deionized (DI) water and stirring at 220 rpm for at least 24 hours to ensure homogeneity. The concentrations were chosen within the entangled regime, determined using the critical concentration 
$c^*$ and entanglement concentration 
$c_e$.  The critical concentration was estimated from the Flory relation 
$c^* = 1/[\eta_i]$, where the intrinsic viscosity 
$[\eta_i]$ was obtained from the Mark–Houwink–Sakurada correlation \cite{tirtaatmadja2006drop}, 
$[\eta_i] = 0.072 M_w^{0.65}$. The entanglement concentration was taken as 
$c_e \approx 6c^*$ \cite{arnolds2010capillary}. All concentrations used in this study, along with their concentration ratios $c/c^*$, are provided in the Supplementary Table \ref{tab:T1}. Shear rheology results are provided in the supplementary material (see Figure \ref{sec:shear_rho}), and relaxation times were determined using capillary breakup and extensional rheometry via dripping-onto-substrate (CaBER-DoS) experiments (see Figure \ref{sec:extensional_rho}).

The coalescence of two identical droplets having a volume of 7.5 $\mu$l was studied in a pendant-pendant configuration, delivered via syringes with Nordson needles having diameter of 1.27 mm attached to syringe pumps. One droplet is advanced towards the other at an extremely low approach velocity, triggering instant coalescence upon contact. The experiment is illuminated from behind using a 45 W LED light source (Nila Zaila) at full intensity. The dynamics are captured with a Photron Fastcam Mini AX-100 high-speed camera operating at 45,000 frames per second. A Navitar zoom lens is employed, with a shutter speed of 1/45,000 s and a resolution of 128×272 pixels. Extracted frames are further processed in MATLAB with a custom-written algorithm. 

\textbf{Numerical Implementation}: We performed two-dimensional simulations of polymeric droplet coalescence using using an open-source framework OpenFOAM~\cite{jasak2007openfoam} integrated with the viscoelastic flow solver \textsc{RheoTool}~\cite{pimenta2017stabilization} to resolve the temporal evolution of the neck radius between droplets. The governing equations include incompressible continuity, momentum conservation with surface tension modeled by the continuum surface force method, and volume-of-fluid interface tracking. Viscoelastic stresses in the droplets were described by the exponential Phan-Thien--Tanner (ePTT) model~\cite{thien1977new,10.1122/1.549481},
\begin{align}
f \, \overset{\nabla}{\boldsymbol{\tau}} + \lambda \, \boldsymbol{\tau} = \eta_p \left( \nabla \mathbf{u} + (\nabla \mathbf{u})^T \right),  ~~\text{and}~~ f = \exp \left( \frac{\epsilon \, \lambda}{\eta_p} \, \mathrm{tr}(\boldsymbol{\tau}) \right),
\end{align}
where $\lambda$ is the polymer relaxation time, $\eta_p$ the polymeric viscosity, $\eta=\eta_s+\eta_p$ the total solution viscosity, and $\epsilon$ an extensibility parameter. Simulations reproduce experimental coalescence dynamics and reveal that increasing polymer concentration enhances elasticity and viscosity, thereby altering the neck growth and reducing the power-law exponent compared to Newtonian droplets.

\section*{Conclusion}

This study presents detailed insights into droplet coalescence in a pendant–pendant configuration. By investigating Newtonian and polymeric droplets across a range of $c/c^*$ values, we identify regime transitions, characterize the associated power-law exponents, and quantify axial curvature. At $c/c^* < 28$, only a single regime is observed, though the exponent $b$ remains below unity, reflecting the dominance of elastic effects. At higher $c/c^*$, two distinct regimes emerge: an initial elasticity-dominated stage, followed by an elasto-viscous regime beyond $t/\lambda \approx 0.05$. In this later stage, polymer chain relaxation enhances viscosity, making viscous effects dominant. This transition is further supported by dimensionless number analysis. All measured $b$ values for polymer droplets lie below 0.5, placing them in the sub-Newtonian regime, which signifies the onset of elastic behaviour in polymeric droplet coalescence. Additionally, our measurements of axial curvature reveal significant deviations from classical assumptions, with discrepancies becoming more pronounced at higher $c/c^*$. In parallel, we performed two-dimensional numerical simulations using a volume-of-fluid framework with the exponential Phan-Thien–Tanner (ePTT) model. The simulations reproduced Newtonian benchmarks, captured the elasticity-induced neck growth, and closely matched experimental observations across concentrations. This experimental–numerical agreement strengthens our interpretation of the elastic-to-elasto-viscous transition and highlights the suitability of the ePTT model in describing coalescence in complex viscoelastic fluids. Finally, we demonstrated the practical relevance of our findings by extending the analysis to a consumer-grade viscoelastic product, shampoo, which also exhibits two distinct regimes during coalescence. 

\section*{Acknowledgement}

A.K.\ acknowledges partial support received from Anusandhan National Research Foundation (India), grant no. CRG/2022/005381. B.M.\ acknowledges partial support from the Department of Science and Technology (DST), India, through the INSPIRE Faculty Fellowship (Grant No.\ DST/INSPIRE/04/2024/003069).

\section*{Supplementary Information}
\setcounter{figure}{0} 
\renewcommand{\thefigure}{S\arabic{figure}}

\setcounter{table}{0} 
\renewcommand{\thetable}{S\arabic{table}} 

\setcounter{section}{0} 
\renewcommand{\thesection}{S\arabic{section}} 



\begin{table}[H]
  \centering
  \caption{Different fluids (Newtonian and polymeric fluids with various $c/c^*$) used in the present study and their corresponding properties.}
  \begin{tabular}{|c| c| c| c| c|}
    \hline
    $c$ (\% w/v) & $c/c^*$ & $\eta$ (Pa.s) & $\lambda$ (ms)  &$Oh$ \\ \hline
    1   & 14      & 0.36  & 62  &1.34 \\ \hline
    1.35   & 19      & 0.79  & 89  &2.94 \\ \hline
    1.7   & 24      & 2.38  & 110 &8.8 \\ \hline
    2   & 28      & 7.5  & 140  &26.8 \\ \hline
    2.27   & 32      & 8.08  & 156 &30.5 \\ \hline
    2.5   & 35      & 14.75 & 168 &52.4 \\ \hline
    2.8   & 39      & 23.6 & 192 &82.7 \\ \hline
    3   & 42      & 38.3 & 210  &134\\ \hline
    3.2   & 45      & 40.7 & 220 &141\\ \hline
    3.6   & 50      & 62.54 & 265 &219\\ \hline
    4   & 56      & 110.8 & 288 &369 \\ \hline
    DI Water &-  &0.001  &-   &0.003\\ \hline
    Glycerol  &-   &0.927  &-  &3.16\\ \hline
    Honey   &-   & 4.4    &-    &14.1\\ \hline
  \end{tabular}
  \label{tab:T1}
\end{table}

\begin{itemize}
\item{$c$ - concentration of polymer in \%weight of polymer/volume of solvent.} 
\item{$c^*$ - critical concentration of the polymer solution.} 
\item{$\eta$ - zero shear viscosity.}
\item{$\lambda$ - relaxation time of the polymer solution.}
\item{$Oh$ - Ohnesorge Number.}
\end{itemize}

\begin{figure}[h]
\centering
\includegraphics[width=.5\linewidth]{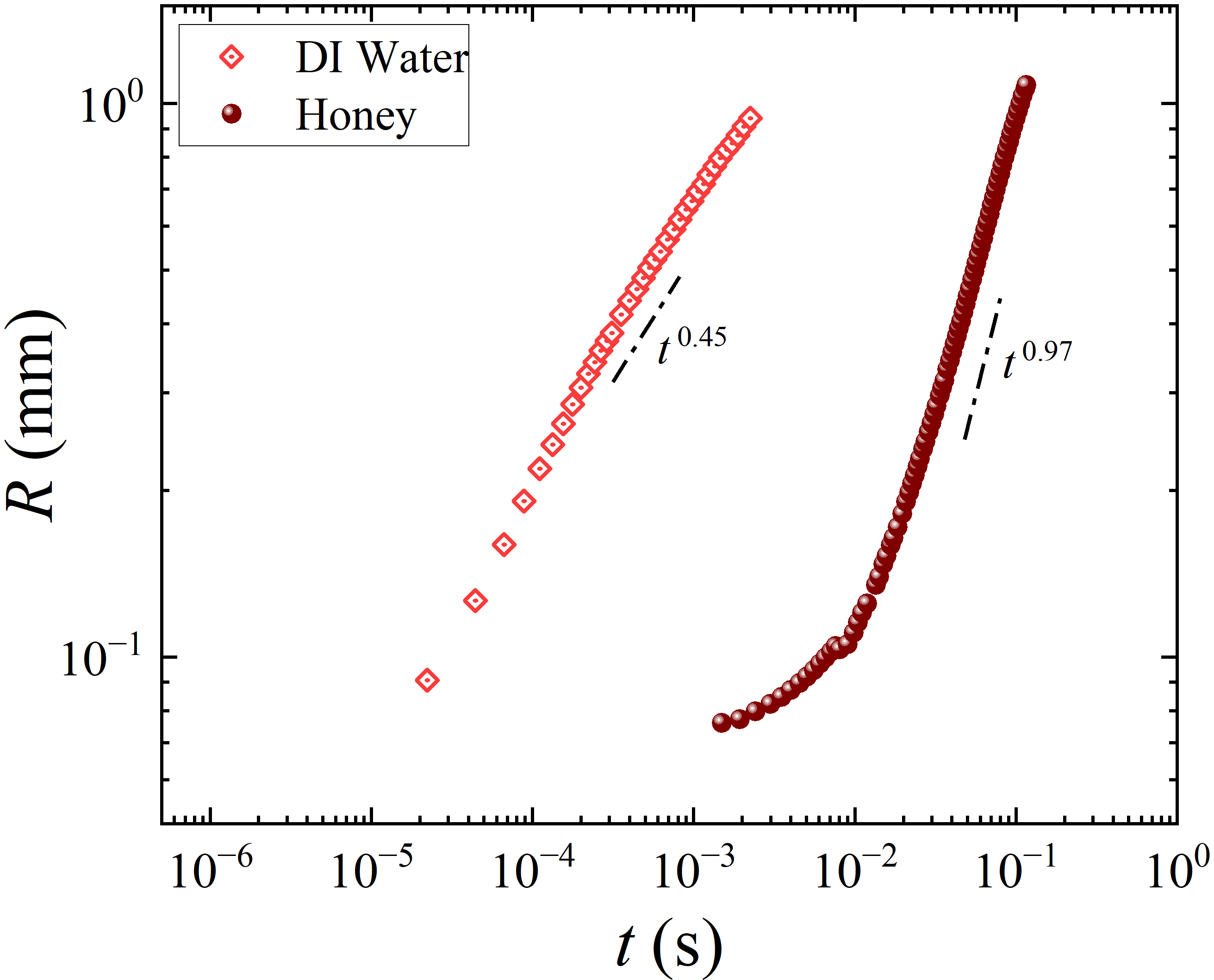}

\caption{Temporal evolution of neck radius for DI water and honey.}
\label{fig:S1}
\end{figure}

Figure \ref{fig:S1} shows the power-law fit to the 
$R$ vs $t$ data for DI water and honey. Water exhibits the value of power-law exponent, 
$b\approx0.45$, consistent with inertial scaling, while the highly viscous honey approaches 
$b\approx 0.97$ as viscous effects dominate.

\begin{table}[h]
 \centering
\caption {Variation of power law exponent ($b$) for Newtonian and polymeric fluids used in the present study.}
 \hspace{0.0cm}   \\
	\begin{minipage}{0.5\textwidth}
	\centering
	\resizebox{\textwidth}{!}{
		\begin{tabular}{|c|c|c|} \hline
 \multirow{2}{*}{Fluids}   &   \multicolumn{2}{c|}{$b$}   \\ \cline{2-3}
              & Regime 1 & Regime 2  \\ \cline{1-3}                     
        DI Water    & 0.45 $\pm$ 0.004  & - \\ \hline
        Glycerol    & 0.71 $\pm$ 0.016  & - \\ \hline
        Honey       & 0.97 $\pm$ 0.02 & - \\ \hline
    $c/c^*$ = 14    & 0.46 $\pm$ 0.01 & - \\ \hline
        19          & 0.42 $\pm$ 0.009 & - \\ \hline
        24          & 0.42 $\pm$ 0.008 & - \\ \hline
        28          & 0.38 $\pm$ 0.008 & - \\ \hline
        32          & 0.36 $\pm$ 0.008 & 0.09 $\pm$ 0.001 \\ \hline
        35          & 0.32 $\pm$ 0.014 & 0.15 $\pm$ 0.008 \\ \hline
        39          & 0.29 $\pm$ 0.008 & 0.16 $\pm$ 0.012 \\ \hline
        42          & 0.28 $\pm$ 0.004 & 0.17 $\pm$ 0.008 \\ \hline
        45          & 0.27 $\pm$ 0.008 & 0.16 $\pm$ 0.004 \\ \hline
        50          & 0.24 $\pm$ 0.012 & 0.14 $\pm$ 0.012 \\ \hline
        56          & 0.24 $\pm$ 0.021 & 0.12 $\pm$ 0.004 \\ \hline
    
\end{tabular}}
	\label{T2}
	\end{minipage}
\end{table}


\newpage
\section {Axial Curvature}
\label{sec:axial_curv}

\begin{figure}[H]
\centering
\includegraphics[width=0.8\linewidth]{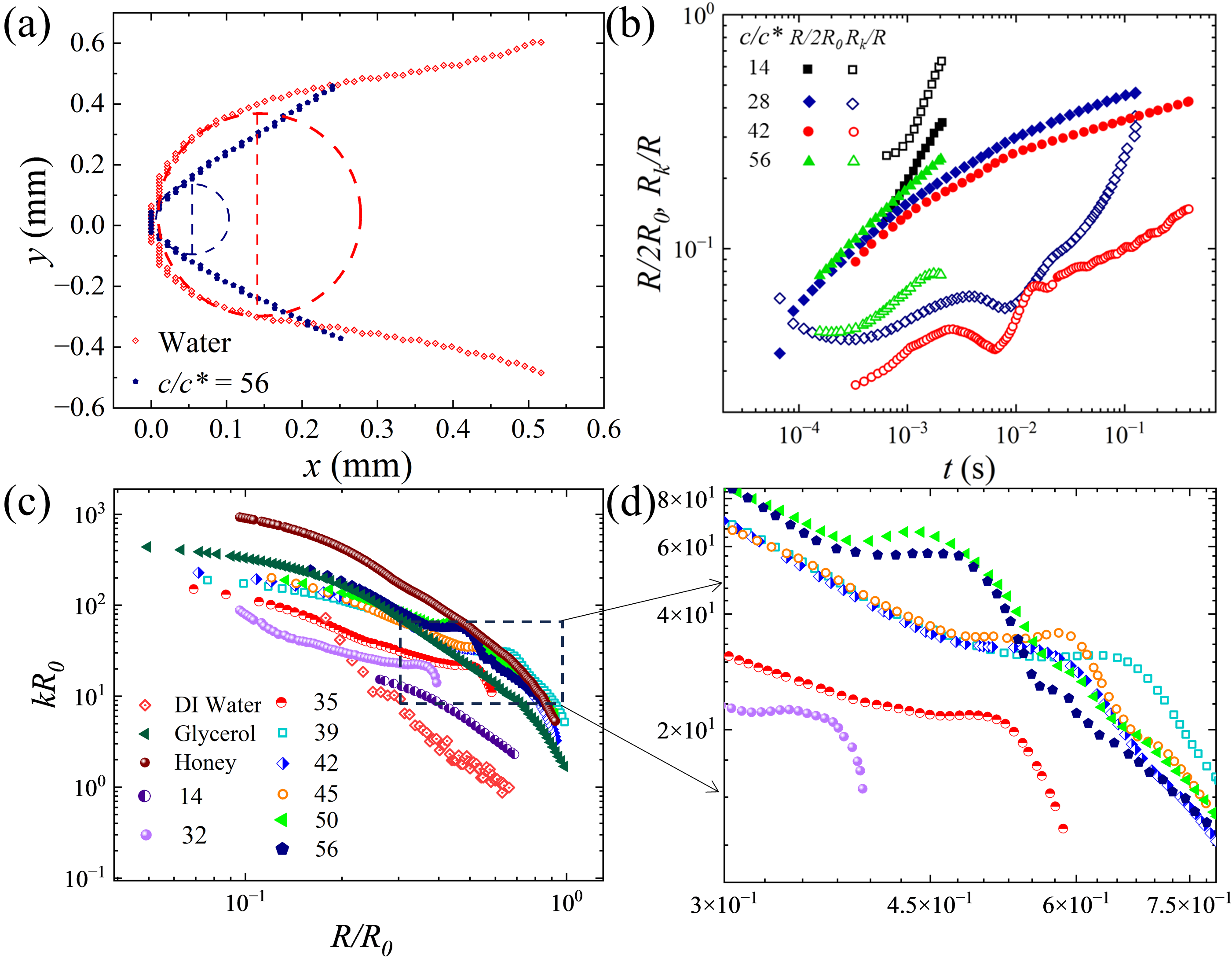}
\caption{(a) Pictorial presentation for the calculation of axial curvature, (b) difference in the $R/2R_0$ and $R_k/R$ for polymer solutions having different $c/c^*$, (c) axial curvature ($k$) normalised by $R_0$ for all fluids examined in the present study, and (d) zoomed-in view of selected cases from figure (c).}
\label{fig:Supl-S2}
\end{figure}

We calculate the axial curvature $k$ experimentally by fitting a 4th-order polynomial to the local profile, approximated as a circular segment.
A pictorial representation of the calculation of 
$k$ is shown in Figure \ref{fig:Supl-S2}(a). The red profile represents DI water, with the vertical line indicating the diameter of the fitted circle ($2R_k$). The blue circle corresponds to the polymer solution with a concentration of  $c/c^*$ = 56. Compared to water, the polymer solution fits a circle with a smaller diameter, indicating a sharper curvature. Next, we compared measured $\tfrac{R_k}{R}$ with the assumed $\tfrac{R}{2R_0}$. Figure \ref{fig:Supl-S2}(b) quantifies this difference, revealing a growing discrepancy between $\tfrac{R_k}{R}$ and $\tfrac{R}{2R_0}$ at higher $c/c^*$, reaching nearly an order of magnitude, demonstrating the breakdown of the simplifying geometrical self-similarity.

We now compare the axial curvature $k$ quantitatively for all test fluids considered in this study, as illustrated in Figure \ref{fig:Supl-S2}(c), at the same normalized neck radius ($R/R_0$). Here, $k$ is normalized by the initial droplet radius $R_0$. As previously discussed, highly viscous Newtonian fluids like honey and glycerol exhibit the highest initial curvature, but their curvature becomes flatter compared to that of more concentrated polymer solutions in the later stage. As discussed above, the viscosity affects axial curvature shapes and coalescence speed as shown by Thoroddsen et al.\cite{thoroddsen2005coalescence}. The sharper curvature of the polymer solutions may be attributed to the stretching of polymer chains during the coalescence process in the narrow region, according to Laplace's law of capillarity~\cite{gennes2004capillarity}. Specifically, the stress within the liquid (relative to atmospheric pressure) is proportional to the axial curvature. The increased curvature seen during the coalescence of polymer solutions results from significant polymer stresses. There are also significant differences in the axial curvature between Newtonian fluids and polymer solutions during the later phases of coalescence, as illustrated in the zoomed plot in Figure \ref{fig:Supl-S2}(d). For polymer solutions with  $c/c^*$ values of 32 and above, we observed a transition in the axial curvature. This transition occurs earlier as $c/c^*$ increases. The transition may be influenced by the increase in elasticity of the polymer solutions as the $c/c^*$ value rises.

\section{Extensional Rheology }
\label{sec:extensional_rho}
\begin{figure}
\centering
\includegraphics[width=.8\linewidth]{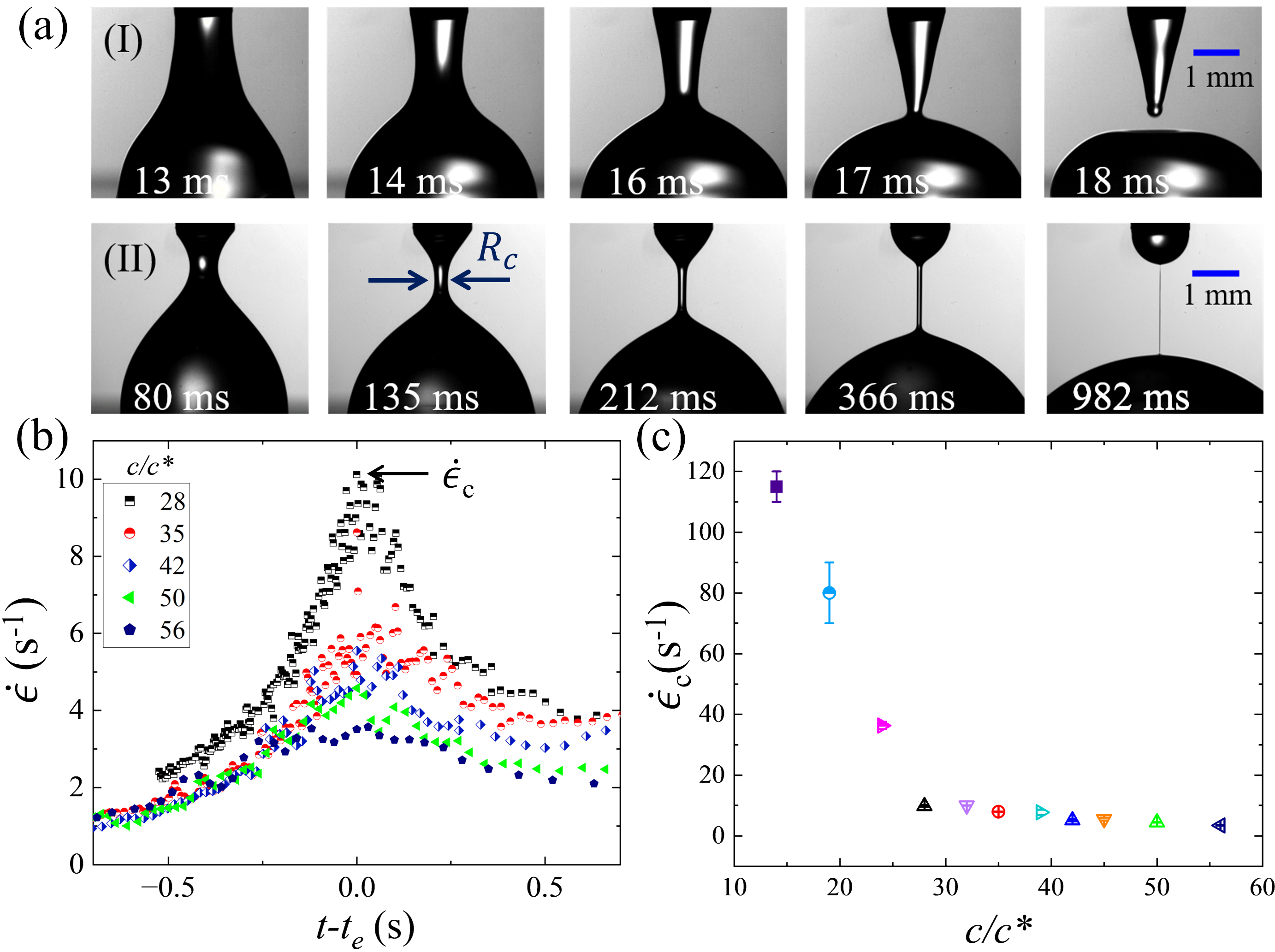}\\
\caption{(a) Snapshots showing capillary pinch-off for DI Water (top row) and PEO having $c/c^*$ = 32 (bottom row), (b) variation of strain rate with time, and (c) critical strain rate for polymer solutions having different $c/c^*$.}
\label{fig:S3}
\end{figure}

 Figure \ref{fig:S3} shows the extensional behavior of the fluids used in this study. Capillary breakup and extensional rheometry dripping-onto-substrate (CaBER-DoS) experiments were performed to determine the relaxation time and extensional strain rate. Fluid was delivered through a 1.27 mm diameter needle attached to a syringe pump, maintaining a flow rate of 15 $\mu$m/min. The needle tip was positioned above a glass substrate, with an optimal height-to-diameter aspect ratio of approximately 3. When the droplet contacts the substrate, it spreads out, forming a capillary between the nozzle tip and the spreading fluid. This process is captured using a high-speed camera operating at frame rates ranging from 1000 to 4000 frames per second, a shutter speed of 1/10,000 s, and a resolution of $1024\times 1024$ pixels. In the case of a DI water (Figure \ref{fig:S3}(a)), the capillary abruptly ruptures after an initial phase of uniform thinning, indicating minimal extensional elasticity. In contrast, for polymers (Figure \ref{fig:S3}(b)), the initial thinning of the bridge is followed by the development of a filament that gradually decreases over time, characterized by a relaxation time $\lambda$, indicating finite extensional elasticity. The relaxation time \cite{amarouchene2001inhibition} of the polymer solution is calculated by fitting an exponential fit as follows $R_c \propto e^{(t-t_e)/3\lambda}$. Here, 
$t$ represents the instant when the droplet touches the substrate, and $t_e$ indicates the start of the elastic regime.

Now, we calculate the elongational strain rate $\dot \epsilon$ for polymer solution using the continuity equation,

\begin{equation}
    \dot \epsilon = \frac{-2}{R_c} \frac{dR_c}{dt}
\end{equation}
 where $R_c$ is the neck radius of droplets in CaBER experiments. Figure \ref{fig:S3}(b) shows that the $\dot \epsilon$ rises to a peak value at the transition before subsequently declining. This point indicates the shift from a Newtonian to a viscoelastic regime. At the transition, this maximum $\dot \epsilon$ is critical strain rate, $\dot \epsilon_c$. The transition from the Inertio-capillary (IC) to the Elasto-capillary (EC) regime occurs when the elongational strain rate ($\dot \epsilon$) in the polymeric liquid reaches a critical value, $\dot \epsilon_c$, at which the extensional stress caused by the background flow of the Newtonian solvent becomes strong enough to induce the coil-stretch transition in the polymer molecule \cite{chandra2024aerodynamic, rajesh2022transition, nguyen2012flexible}. Therefore, if $\dot \epsilon < \dot \epsilon_c$ during a given process involving the polymeric liquid, the behavior will resemble that of a Newtonian liquid, and the elastic properties will have minimal influence. This can be verified by knowing the critical strain rate $\dot \epsilon_c$ for the different polymeric solutions tested in the current study. The quantity $\dot \epsilon_c$ represents the ease with which polymer chains unwind. A smaller value of $\dot \epsilon_c$ indicates that less external deformation is needed to trigger the coil-stretch transition. Essentially, the lower $\dot \epsilon_c$ is, the more easily the polymer chains will elongate under the influence of the surrounding flow. Therefore, for polymer solutions with higher $c/c^*$, the EC regime begins at lower $\dot \epsilon_c$ and it decreases with increasing polymer concentration as illustrated in Figure \ref{fig:S3}(c). However, the reduction in $\dot \epsilon_c$ values is minimal at $c/c^*$ of 28 and higher. 
 
\section{Shear Rheology}
\label{sec:shear_rho}

\begin{figure}[h]
\centering
\includegraphics[width=.8\linewidth]{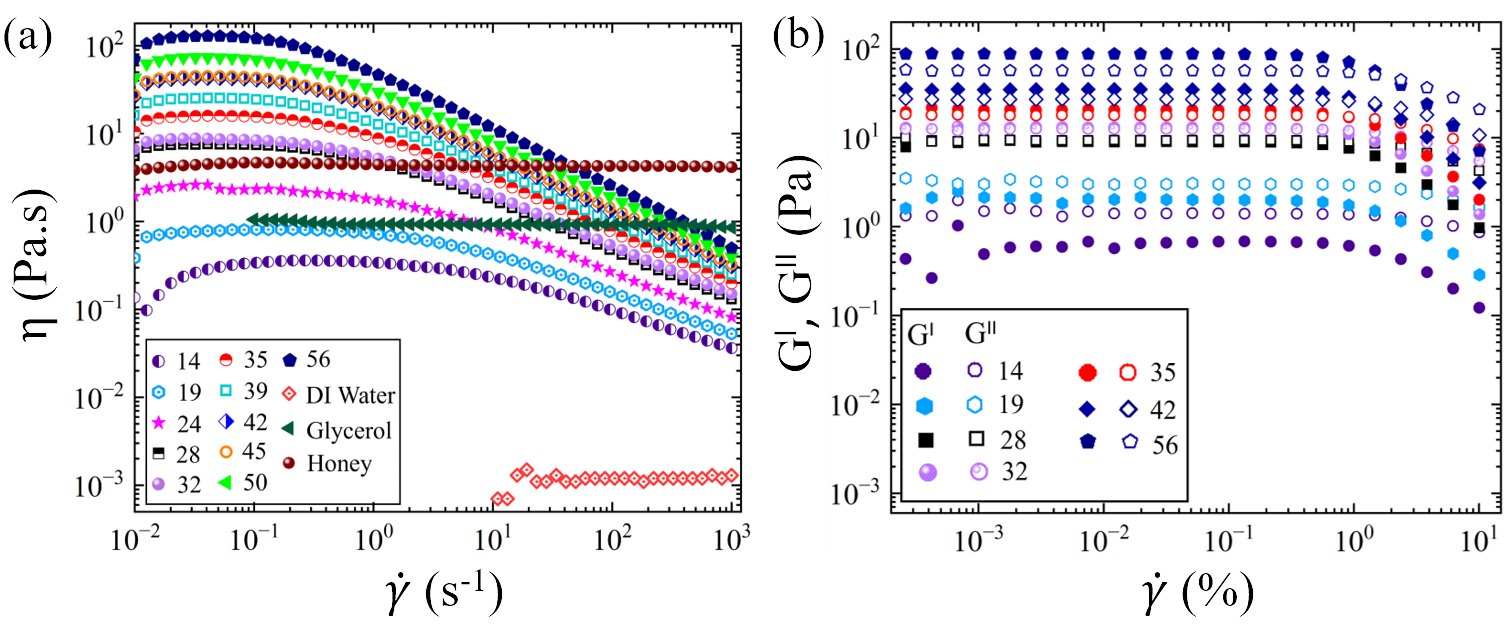}\\
\caption{Variation of viscosity with the shear rate for various Newtonian and polymeric fluids (MCR 302), (b) Amplitude sweep of various polymer solution with different $c/c^*$ to show the degree of viscoelastic behavior of the solutions.}
\label{fig:S4}
\end{figure}

The shear rheology data for the fluids in this study were obtained using an Anton Paar MCR 302 with a cone and plate geometry (40 mm diameter, truncation gap of 80 $\mu$m) at 25 $^\circ$C. We recorded the viscosity variations over a shear rate range from $10^{−1}$ to $10^3$ $s^{−1}$. It was observed that Newtonian fluids exhibit a constant viscosity profile, while polymeric solutions display pronounced shear thinning behavior as shown in Figure \ref{fig:S4}(a). Zero-shear viscosity is presented in Table \ref{tab:T1} for different $c/c^*$ of polymer solution used in the present study. Additionally, we conducted amplitude sweep experiments to gain insight into the viscoelastic behavior of the polymer solutions, as shown in Figure \ref{fig:S4}(b). The storage modulus ($G'$) and loss modulus ($G''$) are represented by filled and open symbols, respectively. At lower $c/c^*$, $G'$ remains smaller than $G''$, indicating that the viscous response dominates over elasticity. At $c/c^* =$ 28, both moduli become nearly equal, and beyond this concentration, $G'$ surpasses $G''$, reflecting the increasing dominance of elastic behavior over viscous behavior.

\bibliographystyle{pnas2009}
\bibliography{apsref}

\end{document}